\documentclass[prb,twocolumn]{revtex4-1} 

\usepackage{amsmath}  
\usepackage{amsfonts} 
\usepackage{graphicx} 

\begin{document}

\title{Getting the Lorentz transformations without requiring an 
invariant speed}

\author{Andrea Pelissetto}
\email{andrea.pelissetto@roma1.infn.it} 

\author{Massimo Testa}
\email{massimo.testa@roma1.infn.it}

\affiliation{Dipartimento di Fisica, Universit\`a degli Studi di Roma ``La Sapienza''\\
   and INFN -- Sezione di Roma I \\
   Piazzale A. Moro 2, I-00185 Roma, Italy}

\date{\today}

\begin{abstract}
The structure of the Lorentz transformations follows purely from the absence of privileged inertial reference frames and the group structure (closure under composition) of the transformations---two assumptions that are simple and physically necessary. The existence of an invariant speed is \textit{not} a necessary assumption, and in fact is a consequence of the principle of relativity (though the finite value of this speed must, of course, be obtained from experiment). Von Ignatowsky derived this result in 1911, but it is still not widely known and is absent from most textbooks. Here we present a completely elementary proof of the result, suitable for use in an introductory course in special relativity. 
\end{abstract}

\maketitle

\section{Introduction}

In standard textbooks, 
the Lorentz transformation equations, which connect inertial reference
frames, are deduced from the invariance of the speed of light,
which implies the invariance of the Minkowski interval. \cite{jackson}
A different approach, however, was followed by von Ignatowsky.\cite{igna} 
Only six years after the formulation of special relativity, he proved 
that the Lorentz transformations arise under quite
general conditions, without assuming \textit{a priori} 
the existence of an invariant speed.  Von Ignatowsky showed that
the only admissible transformations consistent with 
the principle of inertia, 
the isotropy of space, the absence of preferred inertial frames, and 
a group structure (i.e., closure under composition), are the Lorentz transformations,
in which $c$ can be any velocity scale, or the Galilei transformations.
This surprising result shows that the Lorentz transformations are 
not directly related to the 
properties of electromagnetic radiation. Electromagnetism is only 
relevant, if present within the theory, as a way to fix the arbitrary velocity scale, which is then identified with the 
speed of light. This deep and fascinating result, although 
well known in the specialized literature,
\cite{Bacry,berzi,levy1,levy2,Mermin,Torretti,libe1,soneg,bacce}
is not commonly found in textbooks,\cite{note1} 
because its usual proof is rather complicated and uninspiring. It is the purpose of the present paper to present a derivation of the 
von Ignatowsky result using elementary considerations which, in our opinion, shed new light on the result.

\section{Notation for transformations between frames}

We wish to characterize the transformations that relate two different 
inertial frames. 
Let us consider two inertial observers ${\cal O}$ and ${\cal O}'$.
Let ${\bf r} = (x,y,z)$ and $t$ be space and time coordinates for ${\cal O}$
and ${\bf r}' = (x',y',z')$ and $t'$ be the corresponding quantities for 
${\cal O}'$. 

In order to simplify the argument, we will restrict our considerations to the subgroup of transformations involving $x$ and $t$ only, setting $y'=y$ and $z'=z$. This is equivalent to choosing coordinates so that ${\cal O}$ and ${\cal O}'$ are in relative motion along
the $x$ and $x'$ direction.\cite{note2} 

\begin{figure}[h!]
\centering
\includegraphics[width=8.5cm]{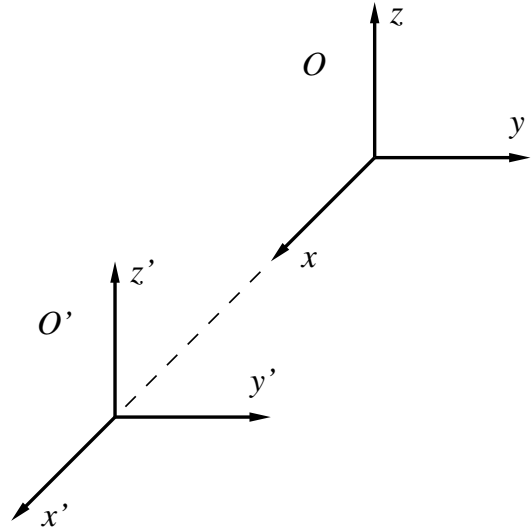}
\caption{Two inertial frames in relative motion along the $x$-axis.}
\label{fig}
\end{figure}

We assume the validity of the principle of inertia: in an inertial
frame free particles undergo rectilinear motion with constant speed.
Therefore, if the trajectory of a particle is 
a straight line in the frame of observer ${\cal O}$ and its speed is constant, 
the trajectory is also a straight line in the frame of observer ${\cal O}'$ and 
also the speed in the new frame is constant. This condition implies \cite{eisenb}
that the two inertial frames
are related by a \textit{linear} transformation. We can therefore write
\begin{eqnarray}
&&{x^0}' = A x^0 + Bx, \label{direct1} \\
&&{x}' = C x^0 + Dx, \label{direct2}
\end{eqnarray}
where we have defined the time coordinates $x^0$ and ${x^0}'$, 
measured in units of distance, as
\begin{eqnarray}
x^0 = {\bar v} t  , \label{vconv}
\end{eqnarray}
where ${\bar v}$ is an \textit{arbitrary} constant with units of speed.
The inverse transformation follows immediately:
\begin{eqnarray}
&&{x^0} = \frac {1} {\Delta} (D {x^0}' - Bx' ), \label{inverse1} \\
&&{x} = \frac {1} {\Delta} (- C {x^0}' + Ax') ,\label{inverse2} 
\end{eqnarray}
with
\begin{eqnarray}
\Delta \equiv AD-BC \neq 0 .
\end{eqnarray}

It is useful to rewrite the transformation in matrix notation,
introducing
\begin{equation}
\Lambda \equiv \begin{pmatrix}
	 A & B \\
	C & D \\
\end{pmatrix}
\end{equation}
and its inverse
\begin{equation}
\Lambda^{-1} = \frac {1} {\Delta}
\begin{pmatrix}
D & -B \\
-C & A \\
\end{pmatrix} .
\end{equation}
Then, we have
\begin{equation}
\begin{pmatrix}
	{x^0}' \\
	x' \\
\end{pmatrix} = \Lambda
\begin{pmatrix}
	x^0 \\
	x \\
\end{pmatrix}  \label{gentrasf}
\end{equation}
and 
\begin{equation}
\begin{pmatrix}
	{x^0} \\
	x \\
\end{pmatrix} = \Lambda^{-1}
\begin{pmatrix}
	{x^0}' \\
	x' \\
\end{pmatrix} . 
\end{equation}

\section{Equivalence of Inertial Reference frames}

The main ingredient in the proof is the requirement that there are no
privileged frames: all inertial frames are equivalent. We wish now to 
express this hypothesis in a more transparent way that allows us to 
put constraints on the matrix $\Lambda$. 

Let us consider two events, $E_1$ and $E_2$, at the same spatial location in frame ${\cal O}$, but separated by a time difference~$\tau$. In ${\cal O}'$ the two events are separated by a time lapse $T'$. 
One natural requirement is that if we consider two different events, $E_3$ and $E_4$, that are at the same spatial location in frame ${\cal O}'$, but separated by the same time difference $\tau$ in ${\cal O}'$, then these two events are separated by a time lapse $T=T'$ in ${\cal O}$. Similarly, we require that if observer ${\cal O}$ measures 
the length $l$ (along the $x$ axis) of a rod that is at rest with respect 
to ${\cal O}'$ and has length $l_0$ in the ${\cal O}'$ frame, 
the same result is obtained by ${\cal O}'$ for an identical rod at 
rest with respect to ${\cal O}$, always along the $x$ axis. These requirements constrain the coefficients of the transformation $\Lambda$.

Let us consider the two events $E_1$ and $E_2$ at the same spatial location in frame ${\cal O}$, separated in time by $\tau$. From Eq.~(\ref{direct1}),
observer ${\cal O}'$ measures a time lapse $T'$ given by 
\begin{eqnarray}
T' =A \tau  .
\end{eqnarray}
Consider now the complementary situation in which an identical clock, at rest 
with respect to ${\cal O}'$, measures the same time interval $\tau$ between two events $E_3$ and $E_4$ that are at the same location in ${\cal O}'$. Correspondingly, from Eq.~(\ref{inverse1}), the observer ${\cal O}$ 
measures a time lapse $T$ given by
\begin{eqnarray}
T =  \frac {D}{\Delta}  \tau  .
\end{eqnarray}
As stated above, if neither ${\cal O}$ nor ${\cal O}'$ is in 
some way privileged, the two time intervals $T$ and $T'$ should be 
identical, i.e., $T'=T$, which implies
\begin{eqnarray}
A = \frac {D} {\Delta}  . \label{tempcov}
\end{eqnarray}

A similar argument applies in the case of a rod 
of rest length $l_0$, oriented along the $x$ axis. 
If the rod is at rest with respect to ${\cal O}'$, Eq.~(\ref{direct2}) implies
that the length $l$ measured by observer ${\cal O}$ satisfies
\begin{eqnarray}
l_0 = D l  .
\end{eqnarray}
Analogously, for a rod at rest in ${\cal O}$, we have, 
from Eq.~(\ref{inverse2}), that the length $l'$ measured by ${\cal O}'$ 
satisfies
\begin{eqnarray}
l_0 = \frac {A} {\Delta} l'  .
\end{eqnarray}
Again, the absence of privileged reference frames requires that $l=l'$ and therefore
\begin{eqnarray}
\frac {A} {\Delta} = D  . \label{spacecov}
\end{eqnarray}

Combining Eqs.~(\ref{tempcov}) and (\ref{spacecov}), we have
\begin{eqnarray}
\frac {A} {\Delta^2} =  A  .
\end{eqnarray}
The solution $A=0$ is not acceptable on physical grounds, 
because it would lead to the meaningless result
${x^0}' = Bx$, so we are forced to choose
\begin{eqnarray}
|\Delta| =1.
\end{eqnarray}
If we restrict ourselves to proper transformations, we have simply
\begin{eqnarray}
\Delta =1  , \label{detone}
\end{eqnarray}
so that
\begin{eqnarray}
A=D  .
\end{eqnarray}
These conditions imply that the transformation relating two different
inertial frames is of the form
\begin{equation}
\Lambda =
\begin{pmatrix}
A & B \\
C & A \\
\end{pmatrix}, \label{relativity1} 
\end{equation}
with
\begin{equation}
A^2 -BC =1, \label{relativity2} 
\end{equation}
the latter condition being a consequence of Eq.~(\ref{detone}).

\section{Group structure}
 
To further constrain the structure of the matrix $\Lambda$, we now add the natural
requirement that 
the transformations connecting two inertial frames constitute a group, i.e., that the combination of two such transformations yields a third transformation of the same form.

We consider two transformations,
\begin{equation}
\Lambda_1 \equiv 
\begin{pmatrix}
A_1 & B_1 \\
C_1 & A_1 \\
\end {pmatrix},
\qquad
\Lambda_2 \equiv 
\begin{pmatrix}
A_2 & B_2 \\
C_2 & A_2 \\
\end {pmatrix}, \label{twoTransformations}
\end{equation} 
and their product,
\begin{eqnarray}
\Lambda_3 &=& \Lambda_1 \Lambda_2 = 
\begin{pmatrix}
A_1 & B_1 \\
C_1 & A_1 \\
\end {pmatrix}
\begin{pmatrix}
A_2 & B_2 \\
C_2 & A_2 \\
\end {pmatrix} 
\nonumber \\ 
&& = 
\begin{pmatrix}
A_1 A_2 +B_1 C_2 & A_1 B_2 +B_1 A_2 \\
C_1 A_2 +A_1 C_2 & A_1A_2 + C_1 B_2 \\
\end {pmatrix}.
\end{eqnarray}
The matrix $\Lambda_3$ is 
an element of the set of admissible matrices only if its diagonal elements are 
equal [see Eq.~(\ref{relativity1})], that is, if
\begin{eqnarray}
B_1 C_2 = C_1 B_2  . \label{grouppro}
\end{eqnarray}

In order to satisfy Eq.~(\ref{grouppro}) for \textit{all} transformations of the form (\ref{twoTransformations}), we have three different possibilities:
\begin{enumerate}
\item[(i)] \label{lorentz}
$B= \alpha  C \label{alpha}$, where $\alpha$ is a nonzero constant;
\item[(ii)] \label{galileo}
$B=0$ and $A=1$, 
where the second equation follows from Eq.~(\ref{relativity2}); or 
\item[(iii)] \label{notinter}
$C=0$ and $A=1$.  
\end{enumerate}

Case (iii) is easily recognized to be physically uninteresting.

Case (ii) corresponds to the Galilean transformations:
\begin{eqnarray}
&&{x^0}' = x^0, \\
&&{x}' = C x^0 + x  ,
\end{eqnarray}
the parameter $C$ being the relative velocity of the two frames in units of 
$\bar{v}$.

In case (i) we change the definition of $x_0$ in 
Eq.~(\ref{vconv}) introducing 
\begin{equation}
\tilde{x}^0 = \bar{c} t = \frac {\bar{c}} {\bar v} x^0
\end{equation}
with 
\begin{eqnarray}
{\bar c} = \frac {\bar v} {\sqrt {|\alpha|}}  .
\end{eqnarray}
Then we obtain 
\begin{equation}
\begin{pmatrix}
	{\hbox{$\tilde{x}$}^0}' \\
	x' \\
\end{pmatrix} = 
\begin{pmatrix}
A & \displaystyle \frac {\alpha}{|\alpha|} C' \\
C' & A \\
\end {pmatrix} 
\begin{pmatrix}
	{\tilde{x}}^0 \\
	x \\
\end{pmatrix},
\end{equation}
where
\begin{eqnarray}
C'= C \sqrt{|\alpha|}.
\end{eqnarray}
Eq.~(\ref{relativity2}) becomes now
\begin{eqnarray}
A^2-\frac {\alpha}{|\alpha|} {C'}^2 =1.
\end{eqnarray}
If $\alpha$ is negative, 
then $\Lambda$ is an orthogonal rotation matrix. In this case, there are transformations
$\Lambda$ such that $\Lambda^n$ is close to the identity for some value of~$n$, and 
this is clearly unphysical.
Therefore, we can exclude
this possibility and we are left with transformations of the form
\begin{equation}
\Lambda = 
\begin{pmatrix}
A & C' \\
C' & A \\
\end {pmatrix},
\end{equation}
where $A^2 - {C'}^2 = 1$. In this case the observer ${\cal O}'$ moves with 
speed $w$ with respect to observer ${\cal O}$, where $w$ is determined by
\begin{equation}
C'/A = w/\bar{c}.
\end{equation}
The condition $A^2 - {C}'^2 = 1$, implies
\begin{eqnarray}
&&A= \gamma=\frac {1}{\sqrt{1-  {w^2} / {{\bar c}^2}}} \\
&&C'= \gamma w / {\bar c}
\end{eqnarray}
so that $\Lambda$ is a generic Lorentz transformation, with the speed of light identified with $\bar{c}$.

\section{Conclusions}

Lorentz and Galilei transformations are the only 
structures compatible with the principles of inertia and 
relativity, i.e., the nonexistence of privileged reference 
frames. This result, which is easily stated, is usually obtained by means
of rather complicated proofs. In this paper we have presented an elementary derivation, suitable for presentation in undergraduate courses in special relativity.


\begin{thebibliography}{00}
\bibitem{note2}
In other words we are considering systems in motion along the $x$ axis: we assume that the most general transformation can be obtained by combining a boost along the $x$ axis with a 
pair of independent orthogonal transformations, i.e., rotations, in the 
two reference frames ${\cal O}$ and ${\cal O}'$.
\end{thebibliography}
\end{document}